\documentclass[useAMS,usenatbib]{mn2e}

\usepackage{graphicx}	
\usepackage{amsmath}	
\usepackage{amssymb}	
\usepackage{multicol}        
\usepackage{bm}		
\usepackage{pdflscape}	
\usepackage{multirow,multicol} 

\def\araa{ARA\&A}%
\def\apj{ApJ}%
\def\aap{A\&A}%
\def\mnras{MNRAS}%
\def\nat{Nature}%
\def\physrep{Phys.~Rep.}%
\title[Limits on turbulent energy propagation]{Limits on turbulent propagation of energy in cool-core clusters of galaxies}
\author[C. Bambic et al.]{C. J. Bambic,$^{1,2}$\thanks{E-mail:
cbambic@umd.edu} C. Pinto,$^{2}$ A. C. Fabian,$^{2}$ J. Sanders$^{3}$ and C. S. Reynolds$^{1,2}$\\
$^{1}$Department of Astronomy, University of Maryland, College Park, MD 20742-2421, USA\\
$^{2}$Institute of Astronomy, Madingley Road, CB3 0HA Cambridge, United Kingdom\\
$^{3}$Max-Planck-Institut f\"ur extraterrestrische  Physik, Giessenbachstrasse 1, 85748 Garching, Germany\\
}
\begin{document}

\date{\today}

\pagerange{\pageref{firstpage}--\pageref{lastpage}} \pubyear{2018}

\maketitle

\label{firstpage}

\begin{abstract}
We place constraints on the propagation velocity of bulk turbulence within the intracluster medium of three clusters and an elliptical galaxy. Using Reflection Grating Spectrometer measurements of turbulent line broadening, we show that for these clusters, the 90\% upper limit on turbulent velocities when accounting for instrumental broadening is too low to propagate energy radially to the cooling radius of the clusters within the required cooling time. In this way, we extend previous Hitomi-based analysis on the Perseus cluster to more clusters, with the intention of applying these results to a future, more extensive catalog. These results constrain models of turbulent heating in AGN feedback by requiring a mechanism which can not only provide sufficient energy to offset radiative cooling, but resupply that energy rapidly enough to balance cooling at each cluster radius. 
\end{abstract}


\section{Introduction}
\label{sec:intro}

The Reflection Grating Spectrometer (RGS) aboard the XMM-\textit{Newton} Observatory has provided increased clarity on the dynamical state of clusters of galaxies. Galaxy clusters, the most massive bound structures in the universe, hold $\sim$ 12\% of their mass in a hot atmosphere of plasma referred to as the intracluster medium (ICM). The low densities ($n \sim 10^{-2} - 10^{-3}$ cm$^{-3}$) and high temperatures ($T \sim 10^7 - 10^8$ K) of this plasma produce short central cooling times, $\sim$ 1 Gyr, yet RGS measurements have revealed a lack of cool gas \citep{Peterson2001}. Some energy source is heating the cluster, maintaining an approximate thermal equilibrium \citep{Peterson2006} and preventing a cooling catastrophe \citep{Fabian1994}.

Feedback from central jetted active galactic nuclei (AGN) provides a source for this energy (see \cite{Fabian2012} and references therein). X-ray observations of nearby clusters have revealed clear cavities (bubbles) inflated by AGN jets \citep{Fabian2000}, with inferred cavity energies which correlate well with radiative losses \citep{Dunn2006}. While the energetics of feedback are well established observationally, the details of how energy from AGN jets is thermalized is an open question. In this letter, we focus on the turbulent heating model of AGN feedback.


Turbulence within the cool core is likely generated by mergers which induce sloshing motions in the ICM \citep{Markevitch2007} as well as buoyancy oscillations (internal waves) driven by the buoyant rise of AGN-inflated bubbles (see Discussion Section 4). These waves are trapped within the cluster core \citep{Balbus1990} where they undergo nonlinear interactions and decay into turbulence, dissipating and heating the cluster. Indeed, if X-ray surface brightness fluctuations are interpreted as turbulent fluctuations, then measurements show that there is enough energy present in bulk turbulence to offset radiative cooling, and this turbulence dissipates rapidly enough to offset measured cooling rates \citep{Zhuravleva2014}. Further measurements from the RGS using both line broadening techniques and resonant scattering measurements have set upper and lower limits respectively on turbulence \citep{Werner2009, Sanders2011, dePlaa2012, Pinto2015}. For many clusters, it is thought that the turbulent energy is sufficient to offset radiative cooling provided the energy is resupplied sufficiently rapidly.

\begin{table*}
\caption{XMM-\textit{Newton}/RGS and Chandra observations used in this paper.}  
\label{table:log}      
\renewcommand{\arraystretch}{1.1}
\small\addtolength{\tabcolsep}{-2pt}

\scalebox{1}{%
\begin{tabular}{ c c c c c}     
\hline  
Source & t\,$^{(a)}$ & XMM Observations & Chandra Observations\\ 
             & (ks) &   &  \\  
\hline
{{NGC 1404}}  & 161.5 & 0781350101 0304940101 & 16231 16232 15233 \\
{{Abell 2204}} & 114.0 & 0306490401 0306490301 0306490201 0306490101 0112230301 & 6104 7940 \\
{{Abell 1835}} & 410.7 & 0551830201 0551830101 0147330201 0098010101 & 6880 6881 7370 \\ 
MACS J2229.7-2755 & 63.3 & 0651240201 & 3286 9374 \\
\hline
\end{tabular}}
\\
$^{(a)}$ RGS net exposure time. XMM observations are used for computing limits on turbulence while Chandra observations are used to compute propagation velocity profiles (see Section 3.3).
\end{table*}

Because the RGS is a slitless spectrometer, the spatial profile of extended sources such as galaxy clusters is convolved with spectral lines, leading to artificial spatial broadening. This artificial broadening means that RGS measurements can only place upper limits on turbulence, rather than provide precise measurements. The most precise measurements of gas motion in clusters come from the Hitomi Soft X-ray Spectrometer (SXS). Hitomi measured the small-scale motions in the Perseus cluster to an unprecedented 10 km/s precision, finding a quiescent velocity dispersion of 164 $\pm$ 10 km/s for the central 60 kpc of the cluster, with a 150 $\pm$ 70 km/s gradient in the velocity across this region \citep{Hitomi2016}. This velocity dispersion corresponds to a turbulent energy of 4\% of the thermal energy, sufficient to offset radiative cooling; however, the low velocity implies that these motions cannot propagate radially throughout the cluster rapidly enough to offset radiative losses at each radius of the cluster \citep{Fabian2017}. Perseus requires a mechanism which can rapidly propagate energy throughout the cluster prior to thermalization. 

In this letter, we show that for the clusters A1835, A2204, and MACS J2229.7-2755, the 90\% upper limit on turbulent velocities is too low to propagate radially to the cooling radius of the clusters within the required cooling time. The problem identified by the low velocity dispersion found by Hitomi for the Perseus cluster is therefore widespread. Thus, even though these clusters may have sufficient energy in bulk turbulence to offset radiative cooling, the radial propagation of this turbulence alone is insufficient to balance radiative losses. 


\section{Data}
\label{sec:data}

The XMM-\textit{Newton} observatory is composed of X-ray imaging cameras and gratings including the European Photon Imaging Cameras (EPIC). Our spectral analysis is performed using the RGS instrument, a slitless spectrometer. We correct for artificial spatial broadening using EPIC/MOS\, 1 surface brightness profiles. Our source extraction and data reduction follow the procedure outlined in \cite{Pinto2015} using the XMM-\textit{Newton} Science Analysis System (SAS) v 16.1.0. 

We performed our background subtraction using a model background produced by the RGS pipeline with normalization based on the count rate measured in CCD\,9. The spectra were converted to SPEX\footnote{www.sron.nl/spex} format through the SPEX task \textit{trafo}. Using MOS\,1 images in the $8-28$\,{\AA} wavelength band, we produce surface brightness profiles. As discussed in Section 1, slitless spectrometers convolve the spatial profile of the source with the observed spectrum. For extended sources such as clusters, this leads to an artificial broadening of the lines at the level,
\begin{equation}
	\Delta \lambda = \frac{0.138}{\mathrm{m}} \Delta \theta \: \mathrm{{\AA}},
\end{equation}
where $\Delta \lambda$ is the wavelength broadening, $\mathrm{m}$ is the spectral order, and $\Delta \theta$ is the source extent in arcseconds. 

\section{Results}
\label{sec:results}
\subsection{Spectral Fits}
We fit our spectra using the SPEX software package, implementing a single-temperature collisional-ionization equilibrium model (\textit{cie}) corrected by galactic absorption (\textit{hot}) and source redshift (\textit{red}). We note that the ICM is a multiphase plasma and thus a single-temperature model is an oversimplification; while using multiple temperature components may improve the overall quality of our fits, these extra components are unlikely to influence the line broadening measurements.

We restrict our spectra to the range $8-28$\,{\AA} for first order spectra, and $8-20$\,{\AA} for second order spectra. This wavelength range maximizes the effective area of RGS and ensures that our source flux is well above the background flux. In order to capture the narrow emission lines characteristic of quiescent clusters, we choose the optimal bin size of 1/3 of the PSF \citep{Kaastra2016}, resulting in lower counts per bin. Thus, we fit our spectra by reducing the C-statistic rather than the Chi-Squared value, ensuring the quality of our fits.

The only free parameters within the \textit{cie} model fit were the temperature, O, Ne, Mg, and Fe metallicities, normalization, and microturbulence velocity $v_{\mathrm{mic}}$, i.e. $\sqrt{2}$ times the standard deviation of the gaussian line fit. The abundance of hydrogen and helium was restricted to solar, and all other metal abundances were coupled to iron. This choice of parameters is similar to \cite{Sanders2013}, with the exception that we do not fit Si because the strongest silicon lines are not redshifted into our spectral band. We fit the redshift for each object, finding values consistent with the measured optical redshifts. The column densities for each object are fixed to values given by the Leiden, Argentine, \& Bonn survey of galactic HI \citep{Kalberla2005}. Figure 1 displays the fits to our spectra. Note that we only show stacked 1st order spectra.

\begin{figure}
   \includegraphics[width=0.98\columnwidth]{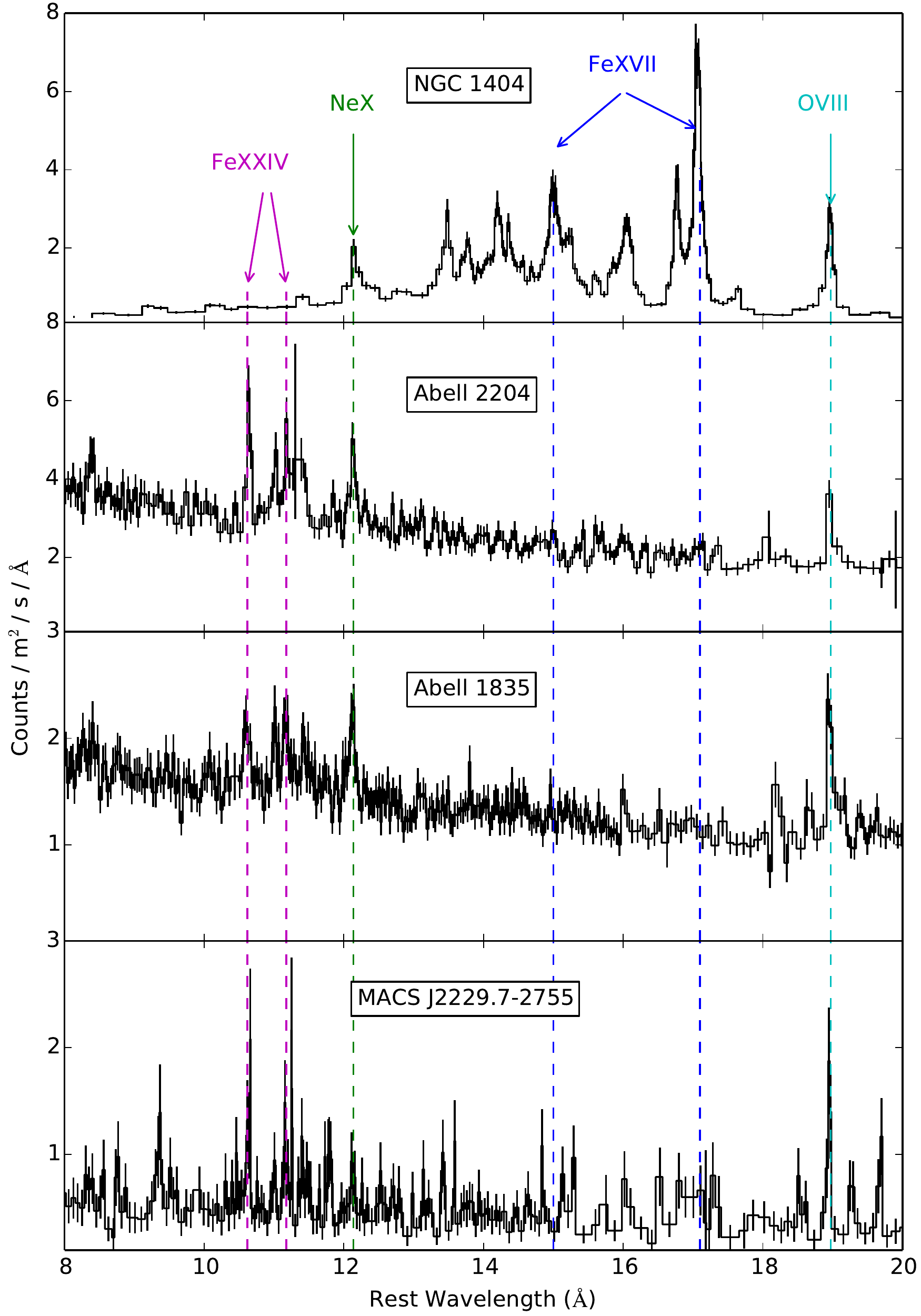}
    \caption{From top to bottom: RGS spectra of NGC 1404 (10), Abell 2204 (10), Abell 1835 (10), and MACS J2229.7-2755 (3). Numbers in parentheses refer to the signal to noise ratio used for plotting. } 
 \end{figure}
 
\begin{table}
\caption{Summary of results.}
\label{tab:results}
\begin{tabular}{lccc}
\hline
Source & Redshift\,$^{(a)}$ &  $<kT>$\,$^{(b)}$ & 90\% Upper Limit \\ 
             &   & (keV) & (km/s) \\  
\hline
{{NGC 1404}}  & 0.006494 & 0.68 $\pm$ 0.01 & 543, 506, 425\\
{{Abell 2204}} & 0.1511 & 3.32 $\pm$ 0.21 & 331, 247, 246\\
{{Abell 1835}} & 0.2532 & 4.14 $\pm$ 0.39  & 223, 224, 244\\ 
MACS J... & 0.324 & 4.38 $\pm$ 0.29   & 207, 209, 216\\
\hline
\end{tabular}
$^{(a)}$ Source redshift (average value taken from the Ned database: https://ned.ipac.caltech.edu/). 
$^{(b)}$ Best-fit temperature for a single isothermal model with no spatial broadening subtracted (see Sect. 3.2). 
90\% upper limits on turbulent velocities from left to right are for no broadening subtracted, 1 gaussian subtracted, and 2 gaussians subtracted respectively (see Sect 3.2). \\ 
\end{table}

For 3 objects (excluding A2204), 2nd order spectra provided the best constraints on turbulent line broadening; 2nd order spectra possess twice the resolution of 1st order spectra. In the case of Abell 2204, we stacked 1st order spectra from multiple observations from both RGS 1 and 2. For A1835, we used the same stacking procedure with second order spectra. We simultaneously fit 1st and 2nd order spectra for MACS J2229.7-2755 and 2nd order spectra for NGC 1404, a nearby (z = 0.0065) bright elliptical galaxy. For simultaneous fits, we coupled the temperatures, metallicities, and redshifts between sectors. 

In order to remove artificial spatial broadening effects, we used the multiplicative \textit{lpro} model. This model convolves an input spatial profile weighted by a parameter $s$ which varies from 0 to 1 with a given spectrum. Previous work freed $s$ on \textit{lpro}, allowing a range between no spatial broadening ($s = 0$) and broadening due to the full extent of the cluster ($s = 1$) to be subtracted from the measured line broadening. Because thermal broadening is accounted for in the \textit{cie} model, any remaining physical broadening can be attributed to turbulence. This method results in a degeneracy between the $v_{\mathrm{mic}}$ and $s$ parameters, a degeneracy which can degrade the statistics for $v_{\mathrm{mic}}$ and drive up the upper limits on turbulence. 

\subsection{Spatial Broadening}

Because the RGS is sensitive to emission from cool gas ($<$ 2 keV), the observed spectra originate from the central 100 kpc, i.e. the ``cool core'' of the cluster. The increased central emission from cooler gas causes the cusp-like structure observed in surface brightness profiles of clusters, while hotter gas causes increased emission in the ``wings'' of the profiles (Sanders et al. 2013). Including the full spatial profile of the cluster in the \textit{lpro} model, i.e. $s = 1$, over-estimates the spatial broadening from cool gas alone and under-estimates the level of turbulence. 

We estimate spatial broadening by first fitting a 3-gaussian model to the spatial profile extracted by \textit{rgsvprof}. The central gaussian represents the coldest core of gas, while the outer gaussian captures the bremsstrahlung continuum from hot hydrogen gas beyond the cluster core. The gaussian between the central and continuum gaussians provides a smooth transition between regions, resulting in a faithful reconstruction of the original profile. Simplifying the model to a 2-gaussian fit tends to over-broaden the central gaussian, leading to an under-estimation of turbulence. Thus, we continue with a 3-gaussian model.

We then produce cumulative spatial profiles identical to those produced by \textit{rgsvprof} from both the central gaussian and central 2 gaussian fits, and include these in the \textit{lpro} model with $s$ set to 1 (or 0.5 for 2nd order spectra). This method conservatively removes artificial broadening due to the spatial extent of the cluster, eliminates degeneracy between the $s$ and $v_{\mathrm{mic}}$ parameters, and prevents under-estimates of turbulent broadening by isolating emission from cooler gas in the cluster center. Figure 2 demonstrates the implementation of this method.

\begin{figure}
   \includegraphics[width=0.97\columnwidth]{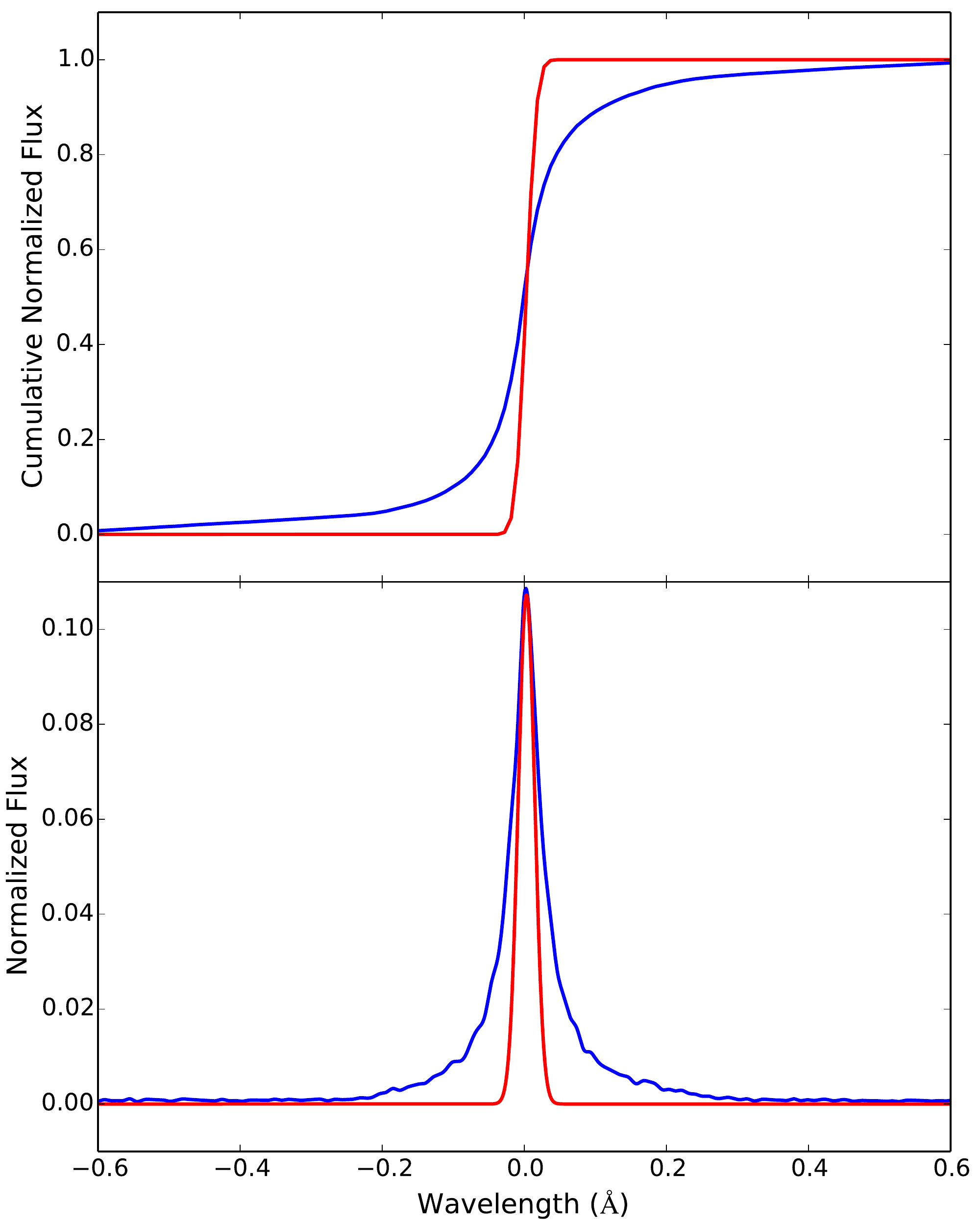}
    \caption{Demonstration of spatial broadening subtraction for NGC 1404. Data is shown in blue, and the fit is shown in red. We fit the sum of 3 gaussian functions to the spatial profile obtained with \textit{rgsvprof}. We then take the central gaussian function to be indicative of the spatial distribution of cool gas (i.e. gas with $T < $ 2 keV) as shown in the bottom panel. This spatial profile is summed to form a normalized cumulative distribution compatible with \textit{lpro} (top panel).}
 \end{figure}

From this point, we were able to compute the turbulent velocity $v_{\mathrm{mic}}$ using an error search, with $\Delta C$ = 2.71 representing the 90\% significance level. The results of our measurements are displayed in Figure 3. Note that all values reported in this paper are 1D turbulent velocities, i.e. $v_{\mathrm{1D}} = v_{\mathrm{mic}}/\sqrt{2}$.

Our results agree with previous work by \cite{Sanders2013}, \cite{Pinto2015}, and \cite{Ogorzalek2017}, who measured turbulent broadening in clusters and elliptical galaxies. \cite{Sanders2013} found a 90\% upper limit of 211 km/s for A1835, consistent with our best limit of 224 km/s for the same object. Furthermore, our best limit on NGC 1404 of 425 km/s (with 2 gaussians subtracted) is fully consistent with the lower limit from resonant scattering measurements by \cite{Ogorzalek2017} of 230$^{+130}_{-90}$ km/s and upper limit from RGS measurements from \cite{Pinto2015} of $\sim$ 700 km/s.

Despite its simplicity, our method is able to reproduce results obtained by far more sophisticated methods, leading us to conclude that our method leads to accurate limits on bulk turbulence in clusters of galaxies.

\begin{figure}
   \includegraphics[width=1.0\columnwidth]{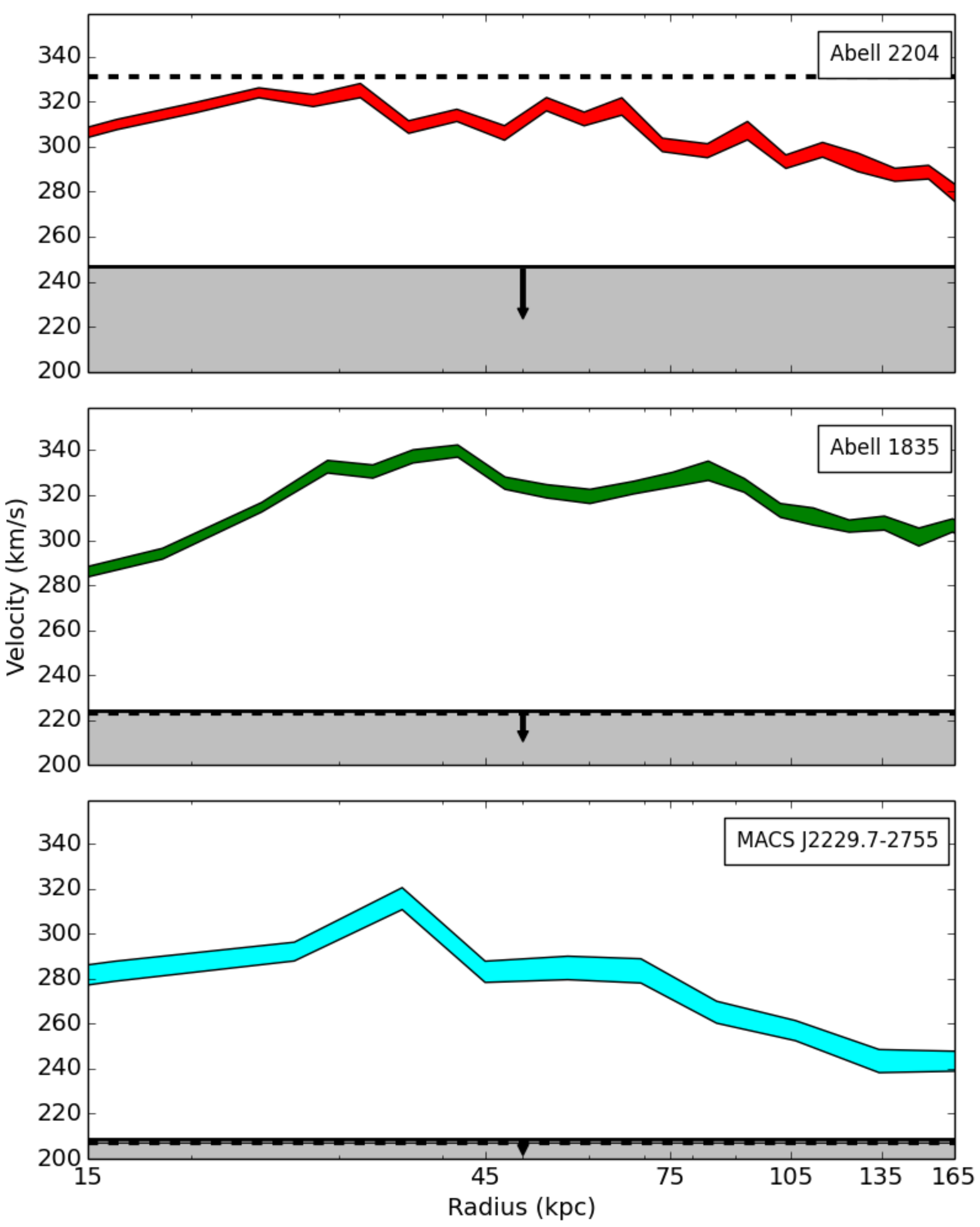}
    \caption{Constraints on turbulent velocities within the three clusters of galaxies. The colored regions represent the 1$\sigma$ error region for the required propagation velocity to balance radiative cooling as a function of radius (Eq. 3). The dashed line represents the 90\% upper limit on 1D turbulent velocities without spatial broadening subtracted, while the solid line shows the 90\% upper limit on turbulent velocities obtained from subtracting spatial broadening due to the central gaussian. Even a conservative subtraction of spatial broadening pushes the upper limit on turbulence below the required value, indicating that within the region of 15 to $\sim$ 165 kpc, bulk turbulence is unable to balance radiative cooling within A2204, A1835, and MACS J2229.7-2755.}
 \end{figure}

\subsection{Propagation Velocity Profiles}

Here, we present a simple formula for the propagation velocity required to offset radiative cooling. In order to balance cooling, the turbulent power $\mathrm{L}_{\mathrm{turb}}$ must equal the cooling luminosity $\mathrm{L}_{\mathrm{cool}}$. If the energy in 3D turbulence ($\frac{3}{2} \mathrm{M}_{\mathrm{gas}} \sigma_{\mathrm{turb}}^2$, where $\sigma_{\mathrm{turb}}$ is the 1D turbulent velocity) is injected over a time $t_{\mathrm{turb}}$, then the power balance results in Equation 2:
\begin{equation}
	t_{\mathrm{turb}} = \frac{3}{2} \frac{\mathrm{M}_{\mathrm{gas}} \: \sigma_{\mathrm{turb}}^2}{\mathrm{L}_{\mathrm{cool}}}.
\end{equation}
Within $t_{\mathrm{turb}}$, turbulence will propagate to a radius $r = \sigma_{\mathrm{turb}} t_{\mathrm{turb}}$. Rearranging this expression to solve for $\sigma_{\mathrm{turb}}$ and converting to dimensionless units where $\mathrm{L}_{\mathrm{cool}} = \mathrm{L}_{44}$ $\times$ 10$^{44}$ ergs/s, $\mathrm{M}_{\mathrm{gas}} = \mathrm{M}_{15}$ $\times$ 10$^{15}$ M\textsubscript{\(\odot\)}, $\mathrm{r} = \mathrm{r}_{\mathrm{kpc}}$ $\times$ 3.09 $\times$ 10$^{21}$ cm, and $\sigma_{\mathrm{turb}} = \sigma_{\mathrm{km/s}}$ $\times$ 10$^5$ km/s, we arrive at Equation 3:
\begin{equation}
	\sigma_{\mathrm{km/s}} = 4.69 \times \Bigg(\frac{\mathrm{r}_{\mathrm{kpc}} \: \mathrm{L}_{44}}{\mathrm{M}_{15}}\Bigg)^{1/3}.
\end{equation}
We note that this method provides a conservative estimate for the required propagation velocity of turbulence, and that using the group velocity of g-modes \citep{Fabian2017} provides an even more stringent upper limit. We used \textit{MBProj2} \citep{Sanders2017} to produce profiles for $\mathrm{L}_{44}$ and ${\mathrm{M}_{15}}$ from Chandra data. The gas mass was estimated from the observed density and the assumption of spherical symmetry. We did not assume a hydrostatic model for fitting the profiles. The 1$\sigma$ error bars on $\sigma_{\mathrm{km/s}}$ (Figure 3) are computed from Monte-Carlo fitting of Equation 3 to the velocity profiles, implementing error chains on each of the measured quantities.

We find that for the three cool-core clusters measured in this study (A2204, A1835, and MACS J2229.7-2755), the best constrained 90\% upper limits on $v_{\mathrm{1D}}$ lie $\sim$ 60 km/s \textit{below} the velocity required to offset radiative cooling. 

\section{Discussion}
\label{sec:discussion}

We have shown that the 90\% upper limits on bulk turbulence within 3 clusters of galaxies are insufficient to propagate energy rapidly enough throughout the cluster to balance radiative cooling. In this argument, we have assumed that turbulent motions originate at the core of the cluster. Our results appear independent of this assumption since if we assume turbulence is driven at 15 kpc (typical radii for cavities in clusters) from the cluster core, the turbulent propagation speed is still too slow to balance radiative cooling.

In addition, we have assumed that the propagation velocity of turbulent motions is constant on timescales comparable to the cooling time. This assumption is testable with a larger sample.

Within the turbulent heating picture of feedback, there is a great deal of debate over how turbulence is generated by the bubbles within the cluster. In some models, turbulence is generated in the wake of the bubble; however, this turbulence remains local to the bubble and does not fill the volume of the cool core. Energy can be propagated by internal waves (g-modes) which are trapped by the entropy gradient of the cluster before interacting and decaying to turbulence; however, the radial group velocity of these waves is too slow to reach the cooling radius of the cluster \citep{Fabian2017}. Furthermore, theoretical/ numerical studies have found the driving of turbulence by AGN jets to be inefficient, arguing instead that bulk motions can be interpreted as weak turbulence and powerful sound waves \citep{RBS15, Yang2016, Weinberger2017, Bourne2017}. These simulations often suffer from the challenge of preserving radio bubbles, a challenge thought to be overcome by magnetic draping \citep{Dursi2008}; however, recent work by \cite{Bambic2018a} argues that magnetic fields are not only ineffective at preserving bubbles through draping, but these same fields \textit{suppress} AGN-driven turbulence. 

In the absence of a clear mechanism for generating and rapidly distributing turbulent energy, it is worthwhile to consider other mechanisms for thermalizing jet energy. Theoretical work with supersonically expanding bubbles tends to inject most kinetic energy into compressive waves or ``sound waves.'' We note that sound waves do not suffer from the limitation discussed in this letter; however, more work within the plasma astrophysics community is required to elucidate dissipation mechanisms which can act in the weakly magnetized, weakly collisional ICM. 


\section{Conclusions}
\label{sec:conclusion}

In this letter, we used XMM-\textit{Newton} RGS observations of three cool-core clusters and one elliptical galaxy to place constraints on the propagation velocity of turbulent motions. Using a simple modeling of the surface brightness profiles of these clusters, we conservatively removed artificial line broadening caused by the spatial extent of our sources. This technique allowed us to measure tight 90\% upper limits on turbulent velocities, consistent with those found by previous works. We find that for three of the cool-core clusters measured in this study, the best constrained 90\% upper limits on the turbulent velocity $v_{\mathrm{1D}}$ lie $\sim$ 60 km/s below the velocity required to offset radiative cooling. Thus, turbulence is unable to propagate energy throughout these clusters rapidly enough to balance radiative cooling at each radius. In this way, our work extends the analysis of \cite{Fabian2017} on the Perseus Cluster, and lays the groundwork for producing a future catalog (Pinto, Fabian, Sanders, et al. 2018 in prep.) which can apply these techniques to a large sample of clusters. These results constrain models of turbulent heating in AGN feedback by requiring a mechanism which can not only provide sufficient energy to offset radiative cooling, but resupply that energy rapidly enough to balance cooling at each cluster radius. 
\
\section*{Acknowledgments}

This work is based on observations obtained with XMM-\textit{Newton}, an ESA science mission funded by ESA Member States and USA (NASA). CJB acknowledges support from the Stamps Family Charitable Foundation as well as the University of Maryland's CMNS Alumni Association. ACF and CP acknowledge support from ERC Advanced grant 340442. CSR thanks the support of the NSF (grant 1333514) and NASA.
\bibliographystyle{aa}

\label{lastpage}
\end{document}